# Smart Radio Spectrum Management for Cognitive Radio

Partha Pratim Bhattacharya, Ronak Khandelwal, Rishita Gera, Anjali Agarwal

Department of Electronics and Communication Engineering
Faculty of Engineering and Technology
Mody Institute of Technology and Science (Deemed University)
Lakshmangarh, Dist – Sikar, Rajasthan, Pin – 332311, India

`hereispartha@gmail.com`

## *ABSTRACT*

*Today's wireless networks are characterized by fixed spectrum assignment policy. The limited available spectrum and the inefficiency in the spectrum usage necessitate a new communication paradigm to exploit the existing wireless spectrum opportunistically. Cognitive radio is a paradigm for wireless communication in which either a network or a wireless node changes its transmission or reception parameters to communicate efficiently avoiding interference with licensed or unlicensed users. It can capture best available spectrum to meet user communication requirements (spectrum management). In this work, a fuzzy logic based system for spectrum management is proposed where the radio can share unused spectrum depending on parameters like distance, signal strength, node velocity and availability of unused spectrum. The system is simulated and is found to give satisfactory results.*

## 1. INTRODUCTION TO COGNITIVE RADIO

Cognitive radio is a kind of wireless communication system in which either a network or a wireless node changes its transmission or reception parameters to communicate efficiently avoiding interference with licensed or unlicensed users. It was thought of as an ideal goal towards which a software-defined radio platform should evolve to a fully reconfigurable wireless black-box that automatically changes its communication variables in response to network and user demand. It is considered to be an intelligent wireless communication system that is aware of its surrounding environment (i.e., outside world), and uses the methodology of understanding-by-building to learn from the environment and adapt its internal states to statistical variations in the incoming RF stimuli by making corresponding changes in certain operating parameters (e.g., transmit-power, carrier frequency, and modulation strategy) in real-time, with two primary objectives in mind:

· highly reliable communications whenever and wherever needed;
· efficient utilization of the radio spectrum.
The idea of cognitive radio was first presented officially in an article by Joseph Mitola III and Gerald Q. Maguire, Jr in 1999 [1]. It was a novel approach in wireless communications that Mitola later described as: The point in which wireless personal digital assistants (PDAs) and the related networks are sufficiently computationally intelligent about radio resources and related computer-to-computer communications to detect user communications needs as a function of use context, and to provide radio resources and wireless services most appropriate to those needs [2].

It was thought of as an ideal goal towards which a software-defined radio platform should evolve: a fully reconfigurable wireless black-box that automatically changes its communication variables in response to network and user demands.

Regulatory bodies in various countries (including the Federal Communications Commission in the United States, and Ofcom in the United Kingdom) found that most of the radio frequency spectrum was inefficiently utilized [2]. For example, cellular network bands are overloaded in most parts of the world, but amateur radio and paging frequencies are not. Independent studies performed in some countries confirmed that observation [3-6], and concluded that spectrum utilization depends strongly on time and place. Moreover, fixed spectrum allocation prevents rarely used frequencies (those assigned to specific services) from being used by unlicensed users, even when their transmissions would not interfere at all with the assigned service. This was the reason for allowing unlicensed users to utilize licensed bands whenever it would not cause any interference. This paradigm for wireless communication is known as cognitive radio. More specifically, the cognitive radio technology will enable the users to determine which portions of the spectrum is available and detect the presence of licensed users when a user operates in a licensed band (spectrum sensing), (2) select the best available channel (spectrum management), (3) coordinate access to this channel with other users (spectrum sharing), and (4) vacate the channel when a licensed user is detected (spectrum mobility).

Cognitive radios have many advantages where the no. of users is high:
- more efficient use of the spectrum
- ensures connectivity
- they are aware of their surroundings and bandwidth availability and are able to dynamically tune the spectrum usage based on location, nearby radios, time of day
- reduced power consumption
- enabling high priority communications to take precedence if needed
- unlimited internet access.

Depending on the set of parameters taken into account in deciding on transmission and reception changes, cognitive radio can be distinguished into the following types -

i) Full Cognitive Radio ("Mitola radio") in which every possible parameter observable by a wireless node or network is taken into account.
ii) Spectrum Sensing Cognitive Radio in which only the radio frequency spectrum is considered.

Also, depending on the parts of the spectrum available for cognitive radio, we can distinguish:

i) Licensed Band Cognitive Radio in which cognitive radio is capable of using bands assigned to licensed users, apart from unlicensed bands, such as U-NII band or ISM band. The IEEE 802.22 working group is developing a standard for wireless regional area network (WRAN) which will operate in unused television channels.
ii) Unlicensed Band Cognitive Radio which can only utilize unlicensed parts of radio frequency spectrum. One such system is described in the IEEE 802.15 Task group 2 specification, which focuses on the coexistence of IEEE 802.11 and Bluetooth.

CR can be viewed as a combined application of SDR and intelligent signal processing with functional elements of radio flexibility, spectral awareness and the intelligence of decision-making. Such cognitive capability allows rapid adaptability to an available communications channel and is an important feature for public safety radio devices to carry the operational requirements of anytime and anywhere communications in the case of an emergency. Also its adaptability to an optimal communications channel will help not only avoid interference to other users but also to improve spectrum efficiency for public safety communications. Thus, CR can greatly enhance spectrum accessibility without causing interference to others. CR devices can also access, and modify, the signal environment including spectrum estimation procedures, signal formats and location awareness information. Hence, CR technologies will enable new public safety systems to share spectrum and/or permit dynamically available spectrum use with existing legacy devices.

## 1.1. THE TECHNOLOGY

Although cognitive radio was initially thought of as a software-defined radio extension (Full Cognitive Radio), most of the research work is currently focusing on Spectrum Sensing Cognitive Radio, particularly in the TV bands. The essential problem of Spectrum Sensing Cognitive Radio is in designing high quality spectrum sensing devices and algorithms for exchanging spectrum sensing data between nodes. It has been shown that a simple energy detector cannot guarantee the accurate detection of signal presence, calling for more sophisticated spectrum sensing techniques and requiring information about spectrum sensing to be exchanged between nodes regularly. Increasing the number of cooperating sensing nodes decreases the probability of false detection [7].

Filling free radio frequency bands adaptively using OFDMA is a possible approach. Timo A. Weiss and Friedrich K. Jondral of the University of Karlsruhe proposed a Spectrum Pooling system in which free bands sensed by nodes were immediately filled by OFDMA subbands.

Applications of Spectrum Sensing Cognitive Radio include emergency networks and WLAN higher throughput and transmission distance extensions.

## 1.2 MAIN FUNCTIONS

The main functions of Cognitive Radios are [8-10]:

i)  Spectrum Sensing: It refers to detect the unused spectrum and sharing it without harmful interference with other users. It is an important requirement of the Cognitive Radio network to sense spectrum holes, detecting primary users is the most efficient way to detect spectrum holes. Spectrum sensing techniques can be classified into three categories:
    o   Transmitter detection: Cognitive radios must have the capability to determine if a signal from a primary transmitter is locally present in a certain spectrum, there are several approaches proposed:
        ▪ matched filter detection
        ▪ energy detection
    o   Cooperative detection: It refers to spectrum sensing methods where information from multiple Cognitive radio users are incorporated for primary user detection.
    o   Interference based detection.
ii) Spectrum Management: It is the task of capturing the best available spectrum to meet user communication requirements [11]. Cognitive radios should decide on the best spectrum band to meet the Quality of Service requirements over all available spectrum bands, therefore spectrum management functions are required for Cognitive radios, these management functions can be classified as:
- spectrum analysis
- spectrum decision

iii) Spectrum Mobility: It is defined as the process when a cognitive radio user exchanges its frequency of operation. Cognitive radio networks target to use the spectrum in a dynamic manner by allowing the radio terminals to operate in the best available frequency band, maintaining seamless communication requirements during the transition to better spectrum

iv) Spectrum Sharing: It refers to providing the fair spectrum scheduling method, one of the major challenges in open spectrum usage is the spectrum sharing.

## 2. BACKGROUND OF THE PRESENT WORK

So from the previous section is may be seen that the main functions of cognitive radio are spectrum sensing, spectrum management, spectrum mobility and spectrum sharing. With these functions it will be able to utilize radio spectrum efficiently.

With cognitive radio which employs Dynamic Spectrum Access (DSA) all available frequency bands including low frequency TV bands and other vacant frequency bands can be put to efficient use and local area network overloading can be avoided as cognitive radio adapts to unusual situations using flexible spectrum access. This increases data transfer speed. The secondary unlicensed users opportunistically utilize these holes for communication without causing interference to primary users as shown in Fig 1.

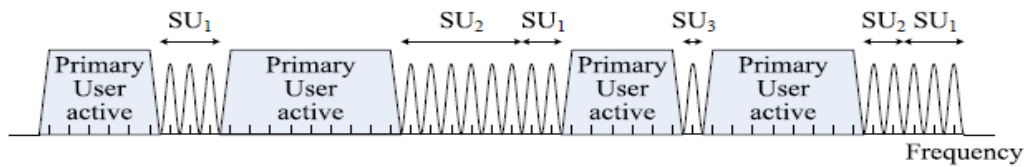

Figure 1: Opportunistic spectrum usage in cognitive radio

The vision is to assign appropriate resources to end users only as long as they are needed for a geographically bounded region, that is, a personal, local, regional, or global cell. The spectrum access is then organized by the network, that is, by the users.

Keeping this scenario in mind a novel fuzzy logic based spectrum management system is proposed in the present work where the radio can utilize spectrum in dynamic manner depending on external parameters.

## 3. PROPOSED FUZZY LOGIC BASED SYSTEM

A fuzzy logic based system for taking decision to use unused spectrum is proposed and studied. Fuzzy logic is used because it is a multi-valued logic and many input parameters can be considered to take the decision. The model of the fuzzy based system is shown in Fig 2.

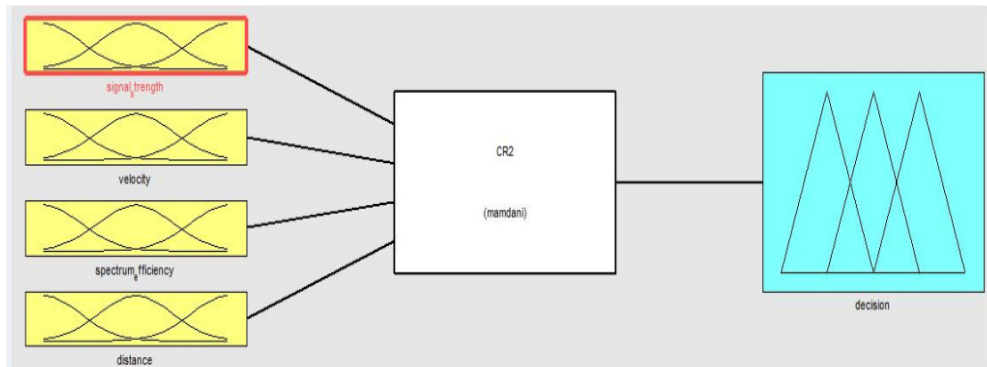

Figure 2. Model of the proposed system

Fuzzy logic is used because it is a multi-valued logic and many input parameters can be considered to take the decision. The signal strength has been considered to be one determining parameter because the radio will decide for change of channel if the signal strength at the intended channel is high. The node velocity is also one input parameter here because more is the velocity more will be the chance for a mobile node to change position and hence quality of service degradation due to low signal strength, particularly at the boundary of service cell or area. The secondary user's velocity is also one input parameter here because more is the velocity more will be the chance for a secondary user to change position and hence possibility of quality of service degradation due to nonavailablity of desired channel. Ratio of the required spectrum by the secondary user to the total available spectrum (spectrum efficiency) has been kept to be the third determining parameter because in this dynamic spectrum access policy radio will use unused vacant spectrum. Unlicensed user should opt for spectrum from the licensed user with maximum vacant spectrum. The distance between the primary licensed and secondary unlicensed user has been considered to be one determining parameter because the secondary user at a closer distance should be given priority to access spectrum from a licensed primary user.

The linguistic variables are kept to be LOW, MEDIUM and HIGH and the membership functions for signal strength, velocity of the secondary user, the ratio of required spectrum by secondary user to total available spectrum and distance between secondary and primary user are shown in Fig 3, 4, 5 and 6 respectively. Gaussian membership functions are used here because with such functions rise time is low and fluctuations are minimum. Based on the knowledge on the linguistic variables 81 IF THEN ELSE fuzzy rules are used to take decision for opportunistic spectrum access (Table 1). At a particular time and place, the unlicensed secondary user with maximum possibility of decision will be allowed to use vacant spectrum.

Mamdani rule is used here and the weight is kept to be 1. Mamdani type fuzzy rule based system (FRBS) provides a natural framework to include expert knowledge in the form of linguistic rules. This knowledge can be easily combined with rules that describe the relation between system input and output. Moreover, Mamdani type FRBS possesses a high degree of freedom to select the most suitable fuzzification and defuzzification interface components as well as the interface method itself. The possibility of spectrum access at a particular location is calculated as
Spectrum access possibility = weight x min value of the membership functions.
A decision value close to 1 is considered to take decision in favor of getting permission for spectrum access. Matlab 7.0 is used for the simulation. The proposed FRBS thus takes decision based on three key parameters according to a predefined rule base. The results are discussed in the next section.

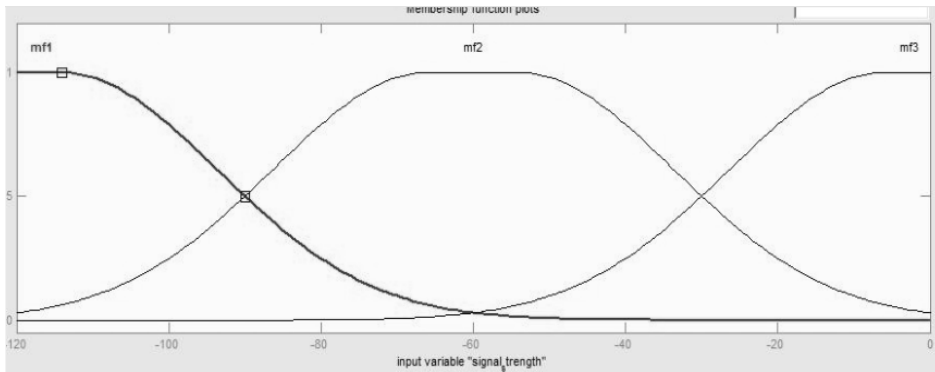

Figure 3. Membership function for signal strength (dBm)

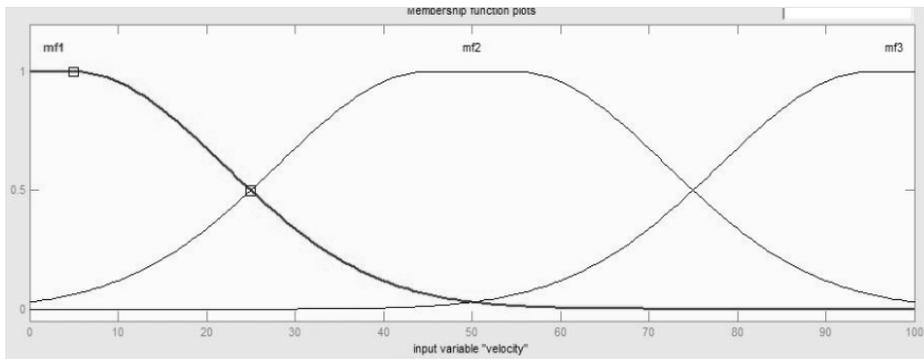

Figure 4. Membership function for node velocity (Km/hr)

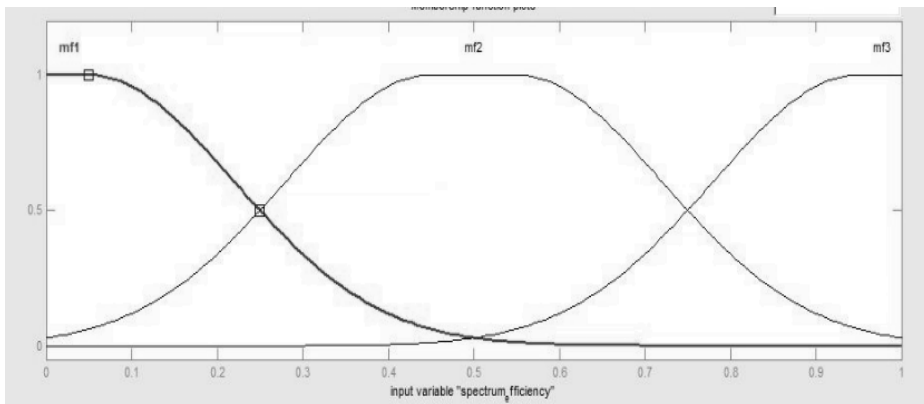

Figure 5. Membership function for ratio of required channels by secondary user to the total number of free channels

International Journal of Distributed and Parallel Systems (IJDPS) Vol.2, No.4, July 2011

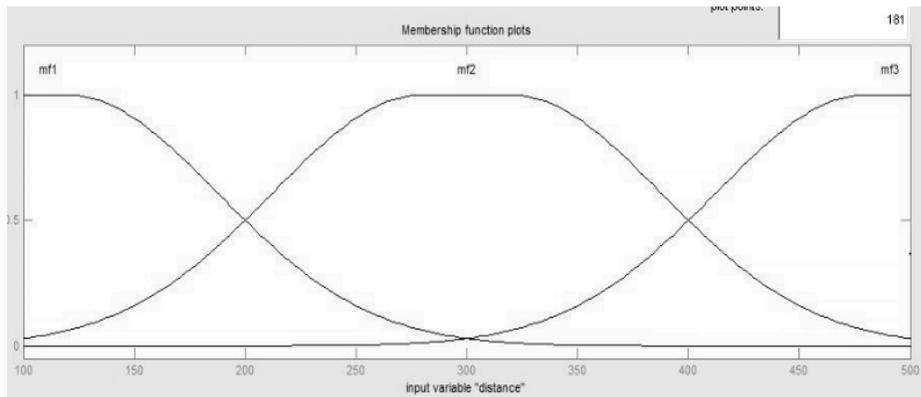

Figure 6. Membership function for distance between primary user and secondary user (meters)

Table1: Rules used for taking the decision

| Sr. no. | Signal Strength | Velocity | Spectrum Efficiency | Distance | Decision |
|---|---|---|---|---|---|
| 1. | Low | Low | Low | Low | High |
| 2. | Low | Low | Low | Medium | High |
| 3. | Low | Low | Low | High | Medium |
| 4. | Low | Low | Medium | Low | High |
| 5. | Low | Low | Medium | Medium | High |
| 6. | Low | Low | Medium | High | Medium |
| 7. | Low | Low | High | Low | Medium |
| 8. | Low | Low | High | Medium | Medium |
| 9. | Low | Low | High | High | Low |
| 10. | Low | Medium | Low | Low | High |
| 11. | Low | Medium | Low | Medium | High |
| 12. | Low | Medium | Low | High | Low |
| 13. | Low | Medium | Medium | Low | High |
| 14. | Low | Medium | Medium | Medium | High |
| 15. | Low | Medium | Medium | High | Medium |
| 16. | Low | Medium | High | Low | Medium |
| 17. | Low | Medium | High | Medium | Low |
| 18. | Low | Medium | High | High | Low |
| 19. | Low | High | Low | Low | High |
| 20. | Low | High | Low | Medium | High |
| 21. | Low | High | Low | High | High |
| 22. | Low | High | Medium | Low | High |
| 23. | Low | High | Medium | Medium | High |
| 24. | Low | High | Medium | High | High |
| 25. | Low | High | High | Low | High |
| 26. | Low | High | High | Medium | Medium |
| 27. | Low | High | High | High | Medium |
| 28. | Medium | Low | Low | Low | High |
| 29. | Medium | Low | Low | Medium | Medium |
| 30. | Medium | Low | Low | High | Medium |

International Journal of Distributed and Parallel Systems (IJDPS) Vol.2, No.4, July 2011

| | | | | | |
|---|---|---|---|---|---|
| 31. | **Medium** | **Low** | **Medium** | **Low** | **Medium** |
| 32. | **Medium** | **Low** | **Medium** | **Medium** | **Medium** |
| 33. | **Medium** | **Low** | **Medium** | **High** | **Medium** |
| 34. | **Medium** | **Low** | **High** | **Low** | **Medium** |
| 35. | **Medium** | **Low** | **High** | **Medium** | **Medium** |
| 36. | **Medium** | **Low** | **High** | **High** | **Medium** |
| 37. | **Medium** | **Medium** | **Low** | **Low** | **Medium** |
| 38. | **Medium** | **Medium** | **Low** | **Medium** | **Medium** |
| 39. | **Medium** | **Medium** | **Low** | **High** | **Medium** |
| 40. | **Medium** | **Medium** | **Medium** | **Low** | **Medium** |
| 41. | **Medium** | **Medium** | **Medium** | **Medium** | **Medium** |
| 42. | **Medium** | **Medium** | **Medium** | **High** | **Medium** |
| 43. | **Medium** | **Medium** | **High** | **Low** | **Medium** |
| 44. | **Medium** | **Medium** | **High** | **Medium** | **Medium** |
| 45. | **Medium** | **Medium** | **High** | **High** | **Medium** |
| 46. | **Medium** | **High** | **Low** | **Low** | **High** |
| 47. | **Medium** | **High** | **Low** | **Medium** | **High** |
| 48. | **Medium** | **High** | **Low** | **High** | **Medium** |
| 49. | **Medium** | **High** | **Medium** | **Low** | **High** |
| 50. | **Medium** | **High** | **Medium** | **Medium** | **High** |
| 51. | **Medium** | **High** | **Medium** | **High** | **Medium** |
| 52. | **Medium** | **High** | **High** | **Low** | **Medium** |
| 53. | **Medium** | **High** | **High** | **Medium** | **Medium** |
| 54. | **Medium** | **High** | **High** | **High** | **Medium** |
| 55. | **High** | **Low** | **Low** | **Low** | **Low** |
| 56. | **High** | **Low** | **Low** | **Medium** | **Low** |
| 57. | **High** | **Low** | **Low** | **High** | **Low** |
| 58. | **High** | **Low** | **Medium** | **Low** | **Low** |
| 59. | **High** | **Low** | **Medium** | **Medium** | **Low** |
| 60. | **High** | **Low** | **Medium** | **High** | **Low** |
| 61. | **High** | **Low** | **High** | **Low** | **Low** |
| 62. | **High** | **Low** | **High** | **Medium** | **Low** |
| 63. | **High** | **Low** | **High** | **High** | **Low** |
| 64. | **High** | **Medium** | **Low** | **Low** | **Low** |
| 65. | **High** | **Medium** | **Low** | **Medium** | **Low** |
| 66. | **High** | **Medium** | **Low** | **High** | **Low** |
| 67. | **High** | **Medium** | **Medium** | **Low** | **Low** |
| 68. | **High** | **Medium** | **Medium** | **Medium** | **Low** |
| 69. | **High** | **Medium** | **Medium** | **High** | **Low** |
| 70. | **High** | **Medium** | **High** | **Low** | **Low** |
| 71. | **High** | **Medium** | **High** | **Medium** | **Low** |
| 72. | **High** | **Medium** | **High** | **High** | **Low** |
| 73. | **High** | **High** | **Low** | **Low** | **Low** |
| 74. | **High** | **High** | **Low** | **Medium** | **Low** |
| 75. | **High** | **High** | **Low** | **High** | **Low** |
| 76. | **High** | **High** | **Medium** | **Low** | **Low** |
| 77. | **High** | **High** | **Medium** | **Medium** | **Low** |
| 78. | **High** | **High** | **Medium** | **High** | **Low** |
| 79. | **High** | **High** | **High** | **Low** | **Low** |

19

| 80. | High | High | High | Medium | Low |
|-----|------|------|------|--------|-----|
| 81. | High | High | High | High | Low |

## 4. RESULTS AND DISCUSSION

The simulation results are shown in Fig 7, 8, 9, 10 and 11. It may be seen from the results that the chance of taking decision increases if the signal strength of the channel offered by primary user is high and the distance between primary and secondary users is low (Fig 7). Similarly, the chance is getting increased when enough free spectrum or channels is available (Fig 8 and Fig 9). Fig 10 shows that as the velocity increases the chance of spectrum accessing is more if the distance is reasonably small. The result is similar for high velocity and low signal strength (Fig 11).

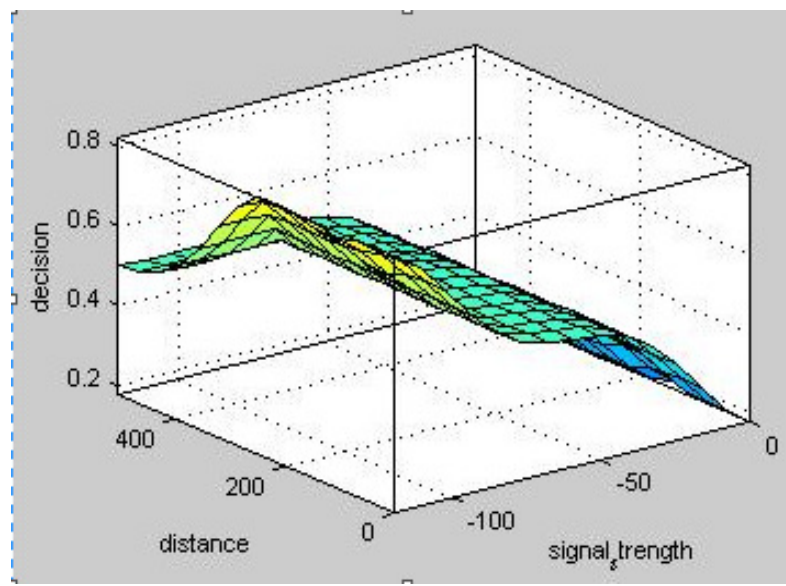

Figure 7. Opportunistic spectrum access decision possibility (Velocity = 50 Km / hr and ratio of required spectrum to available spectrum = 0.5)

International Journal of Distributed and Parallel Systems (IJDPS) Vol.2, No.4, July 2011

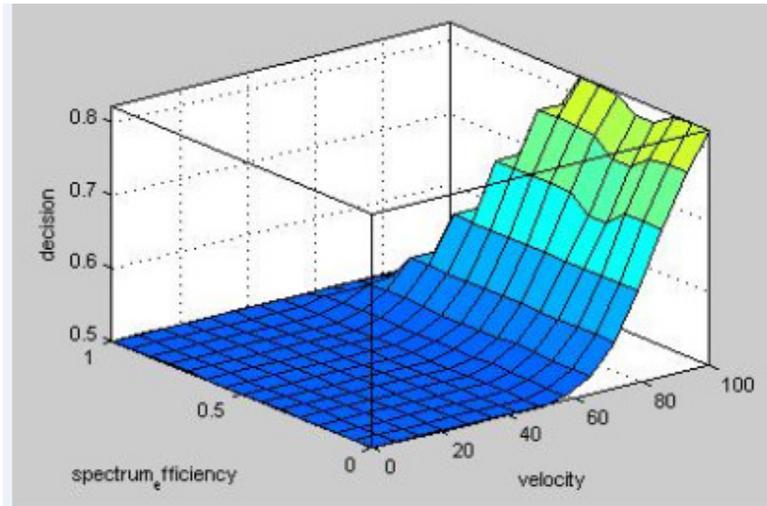

Figure 8. Opportunistic spectrum access decision possibility (Distance between primary and secondary user = 50 meters and signal strength = - 60 dBm)

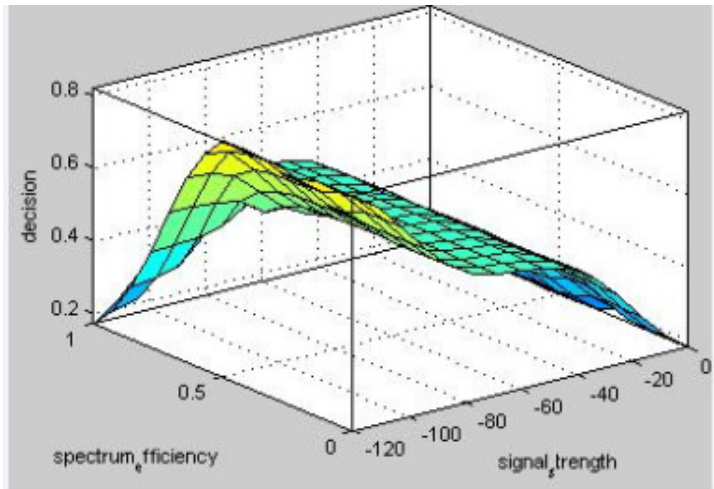

Figure 9. Opportunistic spectrum access decision possibility (Distance between primary and secondary user = 50 meters and user velocity = 50 Km / hr)

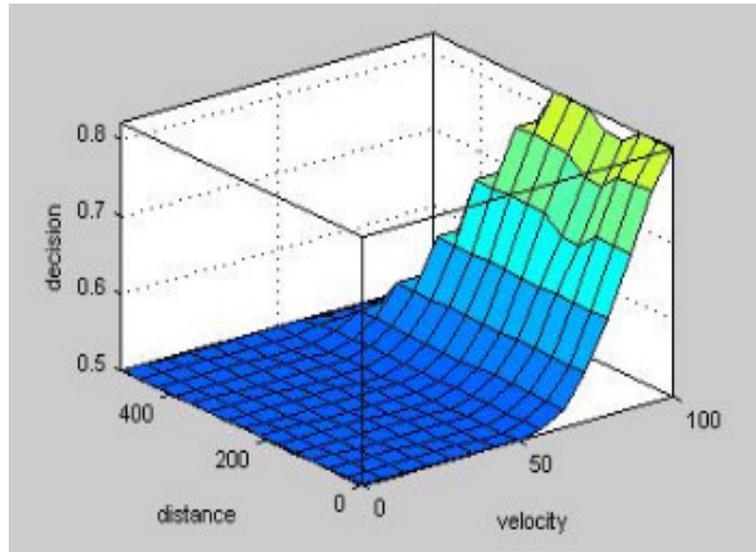

Figure 10. Opportunistic spectrum access decision possibility (ratio of required spectrum to available spectrum = 0.5 and signal strength = - 60 dBm)

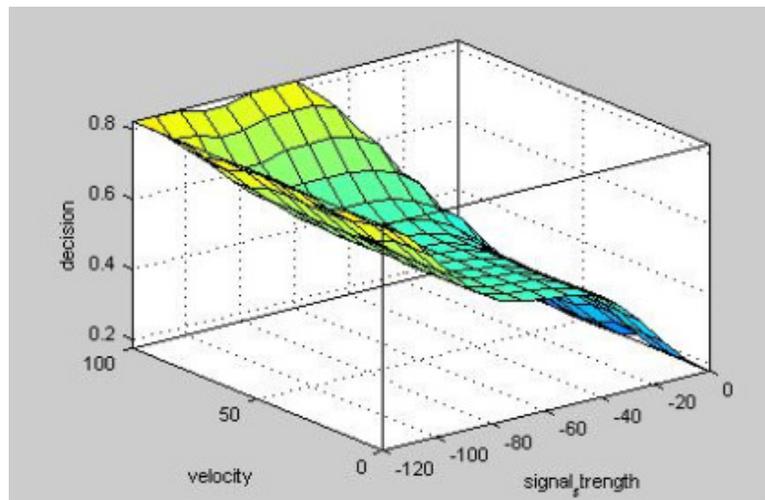

Figure 11. Opportunistic spectrum access decision possibility (distance between primary and secondary users = 50 meters and ratio of required spectrum to available spectrum = 0.5)

So it is clear that the results are desirable and can be used to take decision in practical systems. In present generation systems the mobile node / radio measures the signal strength and the mobile node in future systems would estimate availability of free channels and its velocity using the available standard techniques such as level crossing rate, zero crossing rate etc. [12-16].

## 5. CONCLUSION AND FUTURE SCOPE

Researchers throughout the World are trying to find out the best methods to develop a radio communications system that would be able to fulfill the requirements for a Cognitive radio

22

system. It has been seen that Cognitive radio is the emerging spectrum sharing technology and can be the best option for future generation wireless networks because of present spectrum crisis and uneven use of spectrum. A fuzzy logic based spectrum management technique is proposed here which will help to take wise decision regarding spectrum sharing in cognitive networks. The method considers four input parameters including availability of spectrum and thus is a practicable solution for spectrum sharing. The simulation software programs for the proposed system are neither complex nor consume much time to respond. Hence, it can be easily embedded into application programs and can be implemented in real systems.

**Authors**

Partha Pratim Bhattacharya was born in India on January 3, 1971. He received M. Sc in Electronic Science from Calcutta University, India in 1994, M. Tech from Burdwan University, India in 1997 and Ph.D (Engg) from Jadavpur University, India in 2007.

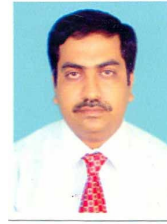

He has 15 years of experience in teaching and research. He served many reputed educational Institutes in India in various positions starting from Lecturer to Professor and Principal. At present he is working as Professor in Department of Electronics and Communication Engineering in the Faculty of Engineering and Technology, Mody Institute of Technology and Science (Deemed University), Rajasthan, India. He worked on microwave devices and systems and mobile cellular communication systems. He has published around 55 papers in refereed journals and conferences. His present research interest includes different aspects of mobile cellular communication and cognitive radio.

Dr. Bhattacharya is a member of The Institution of Electronics and Telecommunication Engineers, India and The Institution of Engineers, India. He received Young Scientist Award from International Union of Radio Science in 2005. He is working as the chief editor, editorial board member and reviewer in many reputed journals.

Ronak Khandelwal was born in India on February 15, 1990. At present she is working in Accenture, Bangalore, India. She passed B.Tech in Electronics and Communication Engineering from Mody Institute of Technology and Science (Deemed University), Rajasthan, India.

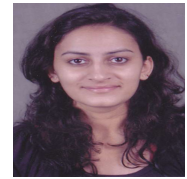

Rishita Gera was born in India on October 5, 1988. She is working in Accenture, Bangalore, India. She passed B.Tech in Electronics and Communication Engineering from Mody Institute of Technology and Science (Deemed University), Rajasthan, India.

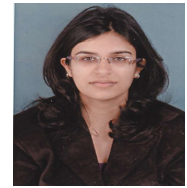

Anjali Agarwal was born in India on January 26, 1988. She is now working in IBM, Noida. She passed B.Tech in Electronics and Communication Engineering from Mody Institute of Technology and Science (Deemed University), Rajasthan, India. She worked at Georgetown University Washington DC 20057 on Sensor networking from June 28 to Aug 2, 2010.

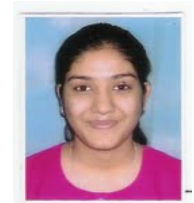